\documentclass[a4paper,fleqn,usenatbib]{mnras}

\usepackage[T1]{fontenc}
\usepackage{ae,aecompl}

\usepackage{graphicx}	
\usepackage{savesym}
\usepackage{amsmath}
\savesymbol{iint}
\usepackage{txfonts}
\restoresymbol{TXF}{iint}
\usepackage{amssymb}	

\title[Four radio quasars at $z>4$: blazars or not?]{VLBI observations of four radio quasars at $z>4$: blazars or not?}

\author[H.-M. Cao et al.]{H.-M.~Cao,$^{1,2}$\thanks{E-mail: hongmin.cao@foxmail.com (HMC)}
S.~Frey,$^{3}$
K.~\'{E}.~Gab\'{a}nyi,$^{3}$
Z.~Paragi,$^{4}$
J.~Yang,$^{5,6}$
D.~Cseh,$^{7}$
\newauthor
X.-Y.~Hong,$^{6}$
and T.~An$^{6,8}$
\\
$^{1}$School of Physics and Electrical Information, Shangqiu Normal University, 298 Wenhua Road, Shangqiu, Henan 476000, China\\
$^{2}$Xinjiang Astronomical Observatory, Chinese Academy of Sciences, 150 Science 1-Street, Urumqi, Xinjiang 830011, China\\
$^{3}$Konkoly Observatory, MTA Research Centre for Astronomy and Earth Sciences, PO Box 67, H-1525 Budapest, Hungary\\
$^{4}$Joint Institute for VLBI ERIC, Postbus 2, 7990 AA Dwingeloo, the Netherlands\\
$^{5}$Department of Earth and Space Sciences, Chalmers University of Technology, Onsala Space Observatory, SE-43992 Onsala, Sweden\\
$^{6}$Shanghai Astronomical Observatory, Chinese Academy of Sciences, 80 Nandan Road, Shanghai 200030, China\\
$^{7}$Department of Astrophysics/IMAPP, Radboud University Nijmegen, PO Box 9010, 6500 GL Nijmegen, the Netherlands\\
$^{8}$Key Laboratory of Radio Astronomy, Chinese Academy of Sciences, Nanjing 210008, China
}

\date{Accepted XXX. Received YYY; in original form 2016 December 9}

\pubyear{2017}

\begin{document}
\label{firstpage}
\pagerange{\pageref{firstpage}--\pageref{lastpage}}
\maketitle

\begin{abstract}
Blazars are active galactic nuclei (AGN) whose relativistic jets point nearly to the line of sight. Their compact radio structure can be imaged with very long baseline interferometry (VLBI) on parsec scales. Blazars at extremely high redshifts provide a unique insight into the AGN phenomena in the early Universe. We observed four radio sources at redshift $z>4$ with the European VLBI Network (EVN) at 1.7 and 5~GHz. These objects were previously classified as blazar candidates based on X-ray observations. One of them, J2134--0419 is firmly confirmed as a blazar with our VLBI observations, due to its relativistically beamed radio emission. Its radio jet extended to $\sim$10~milli-arcsec scale makes this source a promising target for follow-up VLBI observations to reveal any apparent proper motion. Another target, J0839+5112 shows a compact radio structure typical of quasars. There is evidence for flux density variability and its radio ``core'' has a flat spectrum. However, the EVN data suggest that its emission is not Doppler-boosted. The remaining two blazar candidates (J1420+1205 and J2220+0025) show radio properties totally unexpected from radio AGN with small-inclination jet. Their emission extends to arcsec scales and the Doppler factors of the central components are well below 1. Their structures resemble that of double-lobed radio AGN with large inclination to the line of sight. This is in contrast with the blazar-type modeling of their multi-band spectral energy distributions. Our work underlines the importance of high-resolution VLBI imaging in confirming the blazar nature of high-redshift radio sources.
\end{abstract}




\section{Introduction}
 
Active galactic nuclei (AGN) are energetic celestial objects prominent in a wide range of electromagnetic bands, from the radio up to the highest energies (X-rays and $\gamma$-rays). The rich observational phenomena of AGN can be understood in the framework of the orientation-based unified scheme \citep[e.g.][]{Urry1995}. Approximately 10 per cent of the optically-selected AGN are radio-loud \citep{Mushotzky2004}. The radio-loud fraction of AGN may evolve with redshift \citep{Volonteri2011}, but the causes of this evolution are not well understood yet. Radio-loud AGN launch powerful pairs of relativistic jets in the opposite directions perpendicular to the accretion disk which supplies material for the central engine, a supermassive black hole (SMBH, $\sim$$10^6-10^{10}$\,M$_{\odot}$). The powerful relativistic jets are usually found in AGN with high-mass SMBHs, i.e. masses larger than $\sim10^8$\,M$_{\odot}$ \citep[e.g.][]{Sikora2007}. The radio emission of radio-loud AGN originates from incoherent synchrotron processes in the jets \citep[e.g.][]{Blandford1979} and the extended lobes \citep[e.g.][]{Burbidge1956}. If the angle between the approaching jet and the line of sight (i.e. the viewing angle) is small, the radiation from the approaching 
jet is strongly enhanced by relativistic beaming. These AGN are called blazars. Blazars can be divided into two main classes: flat-spectrum radio quasars (FSRQs) and BL Lac objects. The former have prominent optical and ultraviolet (UV) emission lines, while BL Lacs exhibit weak 
(the rest frame equivalent width $\rm{EW}<5$~\AA) or even no emission lines in their spectra \citep[e.g.][]{Fossati1999,Massaro2015}. 
 
FSRQs and BL Lacs are thought to have different accretion mechanisms, i.e. different types of accretion disk 
\citep{Sbarrato2014}. The former are quite luminous in optical, therefore can be seen from large cosmological distances. 
The most distant FSRQ currently known is Q0906+6930\footnote{Very recently, this source is classified as a steep-spectrum radio quasar based on VLBI observations, although at high rest-frame frequencies \citep{Coppejans2016}.} at $z=5.48$ \citep{Romani2006,Zhang2016}, while BL Lacs are rarely found 
at redshift $z>2$ \citep{Plotkin2008}. The multi-band $\rm{log}(\nu \rm{S}_{\nu}) - \rm{log}\,{\nu}$ 
spectral energy distribution (SED) of blazars shows a typical double-hump feature. In the one-zone 
leptonic model \citep{Ghisellini2009} which is widely employed to explain the blazar SED, the low-energy hump 
(peaked at the infrared/X-ray bands) is due to the synchrotron radiation of the relativistic electrons in 
the jet, while the high-energy one (peaked at hard X-ray/$\gamma$-ray bands) is produced by the inverse 
Compton radiation (IC) of the same electron population that causes the synchrontron hump. Depending on whether 
the seed photons are internal or external to the jet, IC can be further classified as Synchrotron-Self-Compton (SSC) 
or External-Compton (EC) emission model. The latter is often used to fit the FSRQ SEDs, 
so as to obtain the physical parameters of the jets, e.g. the jet bulk Lorentz factor $\Gamma$ and  
the viewing angle $\vartheta$ \citep{Ghisellini2009}. High radio loudness \citep[$R \ga 100$, where $R$ is the ratio of rest-frame 5-GHz and 2500-\AA~flux densities, e.g.][]{Sbarrato2013}, high X-ray luminosity ($\nu L_{\nu} \ga 10^{39}$~W; the inverse Compton 
hump moves into the X-ray band) and hard X-ray spectrum are typical characteristics 
of high-redshift blazars \citep[e.g.][]{Sbarrato2013,Ghisellini2015a}. 

Systematic search for high-redshift (here and henceforth, ``high-redshift'' means $z>4$) blazars have been carried 
out by \citet{Sbarrato2012,Sbarrato2013} and \citet{Ghisellini2014}. Their method was to first select highly radio-loud sources (radio loudness parameter
$R>100$) from the Sloan Digital Sky Survey (SDSS) Data Release 7 (DR7) quasar catalogue \citep{Schneider2010} and then to use X-ray data 
to further confirm the blazar identity of the candidates. Blazars could be used to infer their 
parent population: statistically, for one blazar with a viewing angle $\vartheta \le 1/\Gamma$, there should 
be $\sim2\Gamma^2$ radio sources with their jets pointing elsewhere. Besides, once the mass of the 
SMBH powering the blazar is known, the number density of SMBHs hosted 
by radio-loud quasars as a function of redshift could then be explored \citep{Sbarrato2015}.  

Blazars are compact and bright in the radio, and thus ideal targets for very long baseline 
interferometry (VLBI). In particular, high-$z$ blazars allow us to probe the radio emission of the AGN 
in the early Universe. The VLBI images of blazars typically show milli-arcsec (mas) scale compact ``core'' emission or a one-sided core--jet structure. However, the high-$z$ ones rarely have prominent jets. This could be explained by taking into account the relation between the emitted (rest-frame) and observed frequencies in the expanding Universe: $\nu_{\rm em} = (1+z) \, \nu_{\rm obs}$. While the core has typically flat spectrum (with a spectral index $\alpha \ge -0.5$, where $S_{\nu}\varpropto\nu^\alpha$ and $S$ is the flux density), the spectrum of the jet is steep ($\alpha < -0.5$). Therefore, if we observe at a fixed frequency, the high redshift implies a higher emitted frequency and therefore the extended steep-spectrum emission appears fainter relative to the flat-spectrum core \citep{Gurvits1999a}. To date, only one 
high-redshift blazar (J1026+2542, $z=5.27$) is known to have a pronounced jet structure extended to $\sim10$-mas scale which allowed \citet{Frey2015} to estimate the apparent proper motion of the jet components using two-epoch VLBI observations separated by more than 7~yr.

Blazar jets often show apparent superluminal motion, and the apparent proper motion--redshift relation 
can be used to refine the cosmological model \citep[e.g.][]{Vermeulen1994,Kellermann1999,Britzen2008}. Core--jet structures have also been 
proposed as ``standard ruler'' to measure cosmological model parameters using the apparent angular size--redshift relation \citep[e.g.][]{Kellermann1993,Gurvits1994,Gurvits1999b}. Compiling larger samples and incorporating more VLBI observations of high-$z$ blazars are essential for successfully revisiting these cosmological tests because model differences are the most pronounced at the highest redshifts.

High-resolution VLBI imaging is a powerful tool to directly confirm the blazar classification 
of the candidates. If a source is indeed a blazar, it should show Doppler-boosted brightness temperature, flat-spectrum radio core, and possibly rapid and prominent variation in flux density. With this method, \citet{Gabanyi2015} recently confirmed the blazar nature of a newly discovered radio 
quasar, SDSS~J0131$-$0321 with an extremely high luminosity at $z = 5.18$ found by \citet{Yi2014}.

In this paper, we report on our dual-frequency European VLBI Network (EVN) observations of four high-redshift blazar candidates.
Our aims were to confirm their blazar nature and 
to probe their jet structures on $\sim$$1-10$~mas angular scales. If they have jets on these scales, multi-epoch observations could be carried out to measure proper motions. Section~\ref{sec:Target selection} describes our target selection method. The details of the VLBI experiment and the data reduction are given in Sect.~\ref{sec:VLBI observations and data reduction}. We present our results in Sect.~\ref{sec:Results}, and provide the discussion and conclusions in 
Sect.~\ref{sec:Discussion} and Sect.~\ref{sec:Conclusions}, respectively. A flat $\Lambda$CDM cosmological model 
with $H_{0}=71~\rm{km}\,\rm{s}^{-1}\,\rm{Mpc}^{-1}$, $\Omega_{\rm m}=0.27$ and $\Omega_{\Lambda}=0.73$ 
\citep{Spergel2007} is adopted throughout this paper. In this model, 1~mas angular size corresponds to 7.08~pc projected linear size at $z=4$.

\section{Target selection}
\label{sec:Target selection}

\begin{table*}
	\centering
	\caption{Four $z>4$ radio quasars claimed as blazars observed in our EVN experiment.}
	\label{tab:targets}
	\begin{tabular}{ccccc}  
		\hline
		\hline
		Name & Right ascension & Declination & $S_{1.4}$ & $z$ \\
		& h m s & \degr \,\, \arcmin \,\, \arcsec & mJy &\\
		\hline
		J0839+5112  &08 39 46.204 &+51 12 02.88  & $41.6\pm4.0$ &4.390$^\dagger$\\
		J1420+1205  &14 20 48.011 &+12 05 46.40  & $87.3\pm6.6$ &4.034$^\dagger$\\
		J2134$-$0419&21 34 12.012 &$-$04 19 09.67& $311.3\pm18.3$ &4.346\\
		J2220+0025  &22 20 32.608 &+00 25 36.03  & $92.7\pm6.4$ &4.205$^\dagger$\\
		\hline
	\end{tabular}
\\
{\bf Notes.} The equatorial coordinates (J2000) and 1.4-GHz flux densities ($S$) are from the FIRST survey \citep{White1997}. The redshifts ($z$) marked with $\dagger$ are found in \citet{Schneider2010}, the other one is from \citet{Hook2002}.
\end{table*}

We searched in the recent literature \citep[e.g.][]{Volonteri2011,Sbarrato2015,Ghisellini2015b} and found four $z>4$ objects 
which were identified as blazars, have declinations reachable for the EVN but had no published VLBI observations (Table~\ref{tab:targets}). We call these objects blazar candidates. All of them have high radio loudness values ($R>100$), and they were claimed as blazars because of their high X-ray luminosities and hard X-ray spectra, which could not be explained by the AGN corona model. One of our targets (J2220+0025) appeared particularly interesting from the point of view of its extended radio emission on arcsec scale, even before the acquisition of the new high-resolution data. It is somewhat resolved in the 1.4-GHz image taken in the Very Large Array (VLA) Faint Images of the Radio Sky at Twenty-Centimeters (FIRST)\footnote{http://sundog.stsci.edu/} survey \citep{White1997}, with a resolution of about 5~arcsec. Observed at the same frequency but with higher resolution (1.8~arcsec) with the VLA, the source shows an extended ($\sim$10~arcsec) linear structure in the image found in the VLA--SDSS Stripe 82 survey\footnote{http://sundog.stsci.edu/cgi-bin/searchstripe82} \citep{Hodge2011}. However, there is no indication of multiple components in its optical (SDSS) image\footnote{http://www.sdss.org/dr13/}. Therefore it is unlikely that the extended radio structure is caused by gravitational lensing. The remaining 3 sources are unresolved in the FIRST survey images.

\section{VLBI observations and data reduction}
\label{sec:VLBI observations and data reduction}

\subsection{Observations}
\label{subsec:Observations}

Our EVN observations were conducted in e-VLBI mode \citep{Szomoru2008}. The data were taken at 1024~Mbps rate at each participating radio telescope, with two circular polarizations, 8 basebands (or intermediate frequency channels, IF) per polarization and 16~MHz bandwidth per baseband, and transferred to the Joint Institute for VLBI ERIC (JIVE, Dwingeloo, the Netherlands) via optical fiber cables. The data streams were correlated in real time at the EVN software correlator \citep[SFXC,][]{Keimpema2015} with 2~s integration time and 32 spectral channels per IF. The experiment was separated into 4 segments with project codes EC054A, EC054B, EC054C, and EC054D. Three of our targets (J0839+5112, J1420+1205, and J2134$-$0419) were observed with the EVN on 2015 September 15--16 at 1.7~GHz, and on 2015 October 6--7 at 5~GHz. The remaining one (J2220+0025) was first observed on 2016 January 13 at 1.7~GHz, and then on 2016 February 3 at 5~GHz. 

The observations were carried out in phase-referencing mode \citep{Beasley1995}. The radio telescopes were pointed to the a priori positions taken from the FIRST catalogue (Table~\ref{tab:targets}). The nearest suitable phase calibrators, within about $2\degr$ angular separation from the respective targets (see Table~\ref{tab:obs-i}), were selected from the Astrogeo database\footnote{http://astrogeo.org/calib/search.html}. The observing parameters, including the names of radio telescopes participating in the project segments, are listed in Table~\ref{tab:obs-i}.
The bright fringe-finder radio sources used in this experiment were 0528+134 (EC054A), 1156+295 (EC054A and B), DA193 (EC054B), 3C454.3 (EC054C), and CTA102 (EC054D).

The total observing time was about 3~h for each combination of target source and phase calibrator at each frequency. 
The phase-referencing duty circle was 5~min long, with 3.5~min spent on the target source. Therefore, approximately 2~h 
were spent on each target, except for J2134$-$0419. In this case we adopted a different observing strategy, 
because of the high flux density of J2134$-$0419 (see Table~\ref{tab:targets}). To increase the on-target time to 2.5~h, only a couple of phase-referencing scans were distributed over the whole observing period, to allow for the precise determination of the source position. The rest of the time was spent on observing J2134$-$0419 only.

\begin{table*}
	\centering
	\newcommand{\tabincell}[2]{\begin{tabular}{@{}#1@{}}#2\end{tabular}}
	\caption{Observation information of the EVN experiment.} 
	\label{tab:obs-i}
	\begin{tabular}{cccccc}
		\hline
		\hline
		Project segment& \tabincell{c}{Target source(s)} & $\nu_{\rm obs}$ & \tabincell{c}{Participating telescopes} & \tabincell{c}{Phase calibrator}
		&\tabincell{c}{Separation}\\
		&&GHz&&&\tabincell{c}{$^\circ$}\\
		\hline
		EC054A &\tabincell{c}{J0839+5112 \\ J1420+1205 \\ J2134$-$0419}& 1.7 &\tabincell{c}{Ef Hh Jb Mc O8 \\ Tr Wb Sh}
		&\tabincell{c}{J0849+5108 \\ J1415+1320 \\ J2142$-$0437}&\tabincell{c}{1.60 \\ 1.71 \\ 2.12}\\
		\hline
		EC054B &\tabincell{c}{J0839+5112 \\ J1420+1205 \\ J2134$-$0419}& 5   &\tabincell{c}{Hh Jb Mc Nt O8 \\ Tr Wb Sh}
		&\tabincell{c}{J0849+5108 \\ J1415+1320 \\ J2142$-$0437}&\tabincell{c}{1.60 \\ 1.71 \\ 2.12}\\
		\hline
		EC054C &\tabincell{c}{J2220+0025}& 1.7 &\tabincell{c}{Ef Hh Jb Mc O8 \\ Tr Wb}&\tabincell{c}{J2226+0052}
		&\tabincell{c}{1.62}\\
        \hline
		EC054D &\tabincell{c}{J2220+0025}& 5   &\tabincell{c}{Ef Mc Nt O8 Tr \\ Ys Wb Hh}&\tabincell{c}{J2226+0052}
		&\tabincell{c}{1.62}\\
		\hline
	\end{tabular}
\\
{\bf Notes.} Telescope codes: Ef -- Effelsberg (Germany), Hh -- Hartebeesthoek (South Africa), Jb  -- Jodrell Bank Mk2 (United Kingdom), 
Mc -- Medicina (Italy), Nt -- Noto (Italy), O8 -- Onsala (Sweden), Tr -- Toru\'{n} (Poland), Wb -- Westerbork (the Netherlands), 
Ys -- Yebes (Spain), Sh -- Sheshan (China). Hh did not participate in the observations of J0839+5112 because of the high declination of the source.
\end{table*}

\subsection{Data reduction}

The data calibration was conducted in the NRAO Astronomical Image Processing System\footnote{http://www.aips.nrao.edu/index.shtml} 
\citep[{\sc AIPS},][]{Greisen2003}, and generally followed 
the EVN Data Analysis Guide\footnote{http://www.evlbi.org/user\_guide/guide/userguide.html}. A priori 
amplitude calibration was done using the known antenna gain curves and the system temperatures measured at 
the VLBI stations during the observations or nominal system equivalent flux density values. 
Parallactic angle correction for the altitude-azimuth mounted antennas, and ionospheric correction were 
then applied. Manual phase calibration and global fringe-fitting were performed. 
Bandpass correction was carried out using the corresponding fringe-finder source. For the calibration of J2134$-$0419, manual phase calibration was not performed, and the bandpass was corrected using 1156+295 at both 1.7 and 5~GHz. 

The calibrated visibility data for phase calibrator sources were first transferred to and imaged in the {\sc Difmap} program \citep{Shepherd1997}. Antenna-based correction factors were determined for each IF in the first step of the amplitude self-calibration. These were fed back into {\sc AIPS} and applied to the visibilities using the task {\sc clcor}. In this step, only the amplitude corrections exceeding 5 per cent were considered. Global fringe-fitting was then repeated, but the image of each phase calibrator produced earlier in {\sc Difmap} was taken into account this time, to correct for the residual phases resulting from the calibrator's structure. Finally, the phase, delay and delay-rate solutions obtained for the phase-reference calibrators were interpolated and applied to the respective target sources.

In addition to phase-referencing described above, global fringe-fitting was also directly and successfully performed for 
two target sources, J2134$-$0419 and J0839+5112. Due to the lack of the most sensitive EVN antenna (Ef) in the 
project segment EC054B at 5~GHz (see Table~\ref{tab:obs-i}), a longer solution time (10~min) was applied for the direct fringe-fitting on J0839+5112, rather than the 1.5 min used in other cases, in order to properly calibrate the visibility phases on 
the longest baselines to Sh. At last, the calibrated visibility data of each target source were exported from {\sc AIPS}
and loaded into {\sc Difmap} for imaging and fitting brightness distribution models.

The traditional hybrid mapping method (cycles of modeling the brightness distribution and, if allowed by the sufficiently high signal-to-noise ratio, self-calibration of phases and possibly amplitudes) was employed to produce the images of the target sources presented in Sect.~\ref{sec:Results}. The details of the procedure varied from source to source, depending on their properties. For J0839+5112 and J2134$-$0419, we used several cycles of {\sc clean} component modeling and phase (later also amplitude) self-calibration. In these cases we could work with two types of data sets: the one obtained from phase-referencing, and another from direct fringe-fitting. We also fitted Gaussian brightness distribution models to the self-calibrated data in the visibility domain in {\sc Difmap}, to allow for a quantitative description of the source structures by obtaining estimates for component sizes, positions and flux densities.

For the two weaker sources (J1420+1205 and J2220+0025), only phase-referenced data sets were available. For producing the images, we used Gaussian modeling instead of {\sc clean} components, which gave better results due to the extended emission regions of these objects. No phase and aplitude self-calibration was attempted for these weak sources, except for J1420+1205 at 1.7~GHz, where it appeared sufficiently strong for phase-only self-calibration over solution intervals of 10~sec, 
and excluding the long baselines to Hh and Sh.

\section{Results}
\label{sec:Results}

\subsection{Phase-referenced positions}

The phase-referenced data sets were used to measure the astrometric positions of the four target sources. It is possible because the coordinates of the reference sources are accurately known in the International Celestial Reference Frame \citep{Fey2015}. The sources 
were first imaged in {\sc Difmap}, and then the {\sc AIPS} verb {\sc maxfit} was used to derive the coordinates of the brightness peak in the images. The positions measured at 1.7 and 5~GHz were consistent with each other within the uncertainties. The more accurate results obtained at 5~GHz are shown in Table~\ref{tab:position}. The errors arise from the positional uncertainty of the phase calibrator, the thermal noise of the interferometer phases, and systematic errors scaled by the target--calibrator separation \citep[see e.g.][]{Pradel2006}. We estimate that the right ascension and declination coordinates given in Table~\ref{tab:position} are accurate within 0.5~mas.

\begin{table}
	\centering
	\caption{The accurate astrometric positions of the four target sources determined from the phase-referenced EVN observations at 5~GHz.}
	\label{tab:position}
	\begin{tabular}{ccc} 
		\hline
		\hline
		Name        & Right ascension &  Declination \\
		            & h m s & \degr \,\, \arcmin \,\, \arcsec \\
		\hline
		J0839+5112  & 08 39 46.21606 & +51 12 02.8257 \\
		J1420+1205  & 14 20 48.00993 & +12 05 45.9807 \\
		J2134$-$0419& 21 34 12.01074 & $-$04 19 09.8610 \\
		J2220+0025  & 22 20 32.50276 & +00 25 37.5107 \\
		\hline
	\end{tabular}
\end{table}

\subsection{Images and brightness distribution models}

The VLBI image and model parameters obtained in {\sc Difmap} are listed in Tables~\ref{tab:image-p} and \ref{tab:model-p}, 
respectively. The errors of the model parameters are estimated according to \citet{Lee2008}, assuming that the errors are stochastic and independent for each estimated parameter of the component \citep{Fomalont1999}. For the flux densities, we assume an additional 5 per cent error added in quadrature, to account for the uncertainties of the VLBI amplitude calibration which is based on antenna system temperatures and gain curves. Also shown in Table~\ref{tab:model-p} are the source parameters and their errors derived from our measurements: brightness temperature, two-point spectral index between our observing frequencies, and monochromatic luminosity of the core component in the source rest frame. Note that the angular resolution at the two frequencies is different which may cause some flux density contribution from extended structure at 1.7~GHz, leading to an artificial steepening of the derived spectrum.

An interferometer can probe source structures
smaller than the synthesized beam, and the minimum resolvable component size could be estimated according to e.g. \citet{Kovalev2005}. In our experiment, all the fitted component sizes reported in Table~\ref{tab:model-p} exceed the minimum resolvable angular size. The images, modelfit results and derived parameters are described below for each individual target source.

\begin{table*}
	\centering
	\caption{VLBI image parameters of the four target sources.}
	\label{tab:image-p}
	\begin{tabular}{ccccccc}
		\hline
		\hline
		Name & $\nu_{\rm obs}$ & Beam size & P.A. & Peak & RMS & Fig. \\
		& GHz& mas\,$\times$\,mas & \degr & mJy\,beam$^{-1}$ & $\mu$Jy\,beam$^{-1}$ & \\
		\hline
		J0839+5112 & 1.7 &15.4 $\times$ 3.2 & 15 & 37.35 &  41 & \ref{fig:0839}a\\
		           & 5   & 3.4 $\times$ 1.3 & 10 & 36.90 &  58 & \ref{fig:0839}b\\
	  J2134$-$0419 & 1.7 & 5.9 $\times$ 3.3 & 76 &163.22 & 363 & \ref{fig:2134}a\\
		           & 5   & 2.5 $\times$ 1.2 & 81 &138.01 & 127 & \ref{fig:2134}b\\	
		J1420+1205 & 1.7 & 3.8 $\times$ 3.2 & 61 & 10.41 &  60 & \ref{fig:1420}a\\
		           & 5   & 1.6 $\times$ 1.2 & 80 &  4.76 &  88 & \ref{fig:1420}b\\
		J2220+0025 & 5   & 8.3 $\times$ 4.4 &141 &  1.92 &  37 & \ref{fig:2220}\\
		\hline
	\end{tabular}
\\
{\bf Notes.} We only present 5-GHz VLBI image for J2220+0025 (see the text for details). Col.~3 -- synthesized beam size (FWHM), Col.~4 -- position angle of the beam major axis, measured from north through east, Col.~5 -- peak brightness, Col.~6 -- RMS brightness, Col.~7 -- figure number.
\end{table*}

\begin{table*}
	\centering
	\caption{Parameters of the four target sources from the EVN measurements.}
	\label{tab:model-p}
	\begin{tabular}{cccccccccc}
		\hline
		\hline
		Name & $\nu_{\rm obs}$ & Comp. & $S_\nu$ & $\Delta\alpha$ & $\Delta\delta$ &  $\theta$ & $T_{\rm b}$ & $\alpha$ & $L_\nu$ \\
		& GHz &  & mJy & mas & mas & mas & 10$^{9}$K & & 10$^{27}$W\,Hz$^{-1}$ \\
		\hline
J0839+5112  & 1.7 & C & 51.3 $\pm$ 3.3  & 0               & 0                & 2.99 $\pm$ 0.10 & 13.9 $\pm$ 1.8 & $-0.2 \pm 0.1$         & 2.61 $\pm 0.71$  \\
            & 5   & C & 41.6 $\pm$ 3.0  & 0               & 0                & 0.60 $\pm$ 0.02 & 30.0 $\pm$ 4.6 &                        & 2.11 $\pm 0.60$  \\
            &     & E &  1.4 $\pm$ 0.4  &  6.1 $\pm$ 0.1  & --0.8 $\pm$ 0.1  & 0.67 $\pm$ 0.16 &                &                        &                  \\
\hline
J2134$-$0419& 1.7 & C &192.6 $\pm$ 15.3 & 0               & 0                & 1.42 $\pm$ 0.07 & 230.2$\pm$20.5 & $-0.2 \pm 0.1$         & 10.18$\pm 3.03$  \\
            &     & E & 44.0 $\pm$ 6.5  & 21.0 $\pm$ 0.3  &   2.4 $\pm$ 0.3  & 3.25 $\pm$ 0.39 &                &                        &                  \\
            & 5   & C &150.4 $\pm$ 9.7  & 0               & 0                & 0.37 $\pm$ 0.01 & 288.4$\pm$36.2 &                        & 7.95 $\pm 2.25$  \\
            &     & E & 27.8 $\pm$ 3.7  & 12.7 $\pm$ 0.3  &   0.9 $\pm$ 0.3  & 5.19 $\pm$ 0.62 &                &                        &                  \\
\hline
J1420+1205  & 1.7 & C & 20.9 $\pm$ 2.1  & 0               & 0                & 3.45 $\pm$ 0.26 & 4.0 $\pm$ 1.0  & $-0.9 \pm 0.2$         & 2.75 $\pm 1.34$  \\
            &     &NW & 27.0 $\pm$ 7.7  &--275.2$\pm$ 4.0 &1299.3 $\pm$ 4.0  & 22.7 $\pm$ 6.4  &                &                        &                  \\
            & 5   & C &  8.0 $\pm$ 1.3  & 0               & 0                & 1.12 $\pm$ 0.15 & 1.6 $\pm$ 0.7  &                        & 1.05 $\pm 0.58$  \\
\hline
J2220+0025  & 1.7 & C &  3.7 $\pm$ 1.3  & 0               & 0                & 11.6 $\pm$ 3.5  & 0.06$\pm$0.05  & $-0.4 \pm 0.5$         & 0.23$^{+0.48} _{-0.16}$ \\
            &     &SE &  4.5 $\pm$ 1.7  &2447.7 $\pm$ 8.0 &--1928.6 $\pm$ 8.0& 24.0 $\pm$ 8.1  &                &                        &                         \\
            & 5   & C &  2.5 $\pm$ 0.5  & 0               & 0                &  2.9 $\pm$ 0.4  & 0.07$\pm$0.03  &                        & 0.15$^{+0.26} _{-0.10}$ \\
 	\hline
	\end{tabular}
\\
{\bf Notes.} Col.~2 -- observing frequency, Col.~3 -- component designation, Col.~4 -- flux density, Cols.~5--6 -- relative right ascension and declination with respect to the core, Col.~7 -- fitted Gaussian component diameter (FWHM), Cols.~8--10 -- rest-frame brightness temperature, spectral index and monochromatic luminosity of the main (core) component. For J2220+0025, the flux density errors are likely overestimated, and the core luminosity upper and lower bounds are derived by assuming the spectral index upper and lower bounds.
\end{table*}

\subsubsection{J0839+5112}
\label{subsec:J0839+5112}

\begin{figure}
\centering
\includegraphics[width=0.40\textwidth]{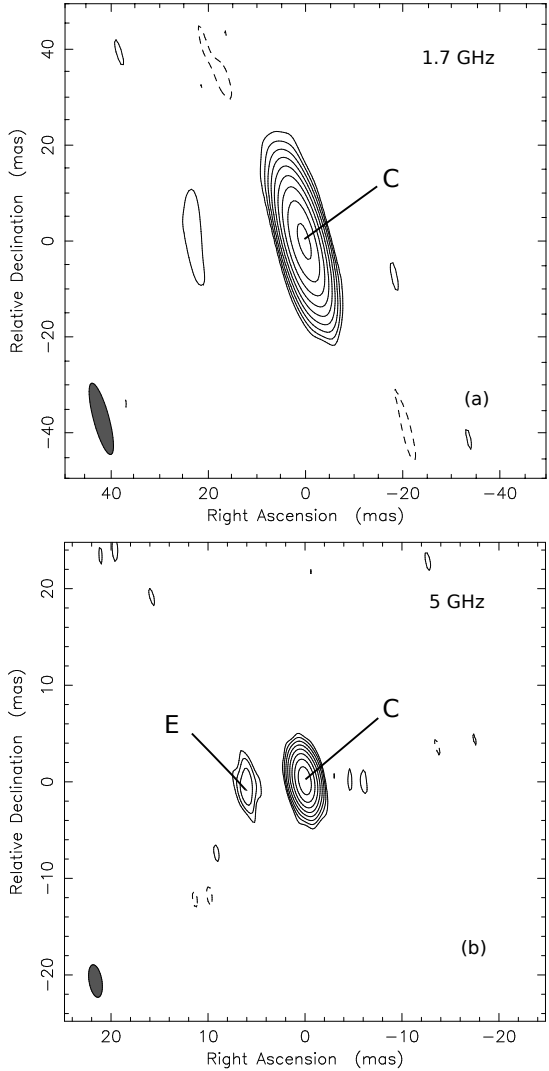}
\caption{Naturally weighted VLBI images of J0839+5112 at 1.7 and 5~GHz. 
The first contours are drawn at $\pm$3$\sigma$ image noise level. 
The positive contours increase by a factor of 2. The synthesized beam (FWHM) is shown at the lower-left corner of each image. The image parameters are listed in Table~\ref{tab:image-p}.}
\label{fig:0839}
\end{figure}

The images made from the phase-referenced and fringe-fitted visibility data sets are practically the same in terms of source structure, peak brightness and noise level for this source at both observing frequencies. 
In Fig.~\ref{fig:0839}, we show the 1.7-GHz image of J0839+5112 produced from the fringe-fitted data set, while at 5~GHz, the phase-referenced image is displayed. The latter is somewhat less noisy than its fringe-fitted counterpart, due to the better calibrated phases on the long baselines from the European antennas to Sheshan made possible by the strong phase-reference calibrator. The relative coordinates in the images are measured with respect to the location of the brightness peak.  

A single component is seen in the 1.7-GHz image, therefore one circular Gaussian model was used to 
fit the source brightness distribution. At 5~GHz, there is a weak jet-like component to 
the east of the bright core at about 6~mas separation (Fig.~\ref{fig:0839}b). In this case, we modeled the source with two circular Gaussian components (Table~\ref{tab:model-p}).

The dual-frequency observations allowed us to calculate the two-point spectral index of the dominant (core) component present in both images,
using the model-fitted flux densities (Table~\ref{tab:model-p}). The value $\alpha=-0.18$ indicates a flat radio spectrum. We also derived the core brightness temperature in the source rest frame (Table~\ref{tab:model-p}) using the formula
\begin{equation}
    T_{\rm{b}}=1.22\times10^{12}(1+z)\frac{S_{\nu}}{\theta^2\nu_{\rm obs}^2}\,\,[{\rm K}]
	\label{eq-2}
\end{equation}
\citep[e.g.][]{Condon1982,Lee2008}, where $z$ is the redshift, $S_{\nu}$ the flux density in Jy, $\nu_{\rm obs}$ the observing frequency 
in GHz, and $\theta$ the full width at half-maximum (FWHM) size of the circular Gaussian model component in mas. The 
monochromatic luminosity of the dominant component in the source rest frame (Table~\ref{tab:model-p}) was calculated using
\begin{equation}
    L_\nu=4 \pi {D^2_{\rm L}} \frac{S_\nu}{(1+z)^{1+\alpha}}
	\label{eq-3}
\end{equation}
\citep{Hogg2002}, where $D_{\rm L}$ is the luminosity distance and $\alpha$ the spectral index. 

The 1.7-GHz VLBI flux density of $51.3\pm3.3$~mJy, although measured with much higher angular resolution, is 23 per cent higher than VLA value at 1.4~GHz in the FIRST survey, $41.6\pm4.0$~mJy (Table~\ref{tab:targets}). Given the very close observing frequencies, this can only be explained by flux density variability between the two measurement epochs. The compact core--jet radio structure (Fig.~\ref{fig:0839}), the flat core spectrum and the implied variability are all characteristic of blazars. On the other hand, the measured brightness temperature, $T_{\rm b} = (3.0\pm0.5) \times 10^{10}$~K (Table~\ref{tab:model-p}) does not indicate Doppler-boosted emission. If we adopt the equipartition brightness temperature \citep[$T_{\rm eq} \approx 5 \times 10^{10}$~K,][]{Readhead1994} as the intrinsic brightness temperature ($T_{\rm int}$), the Doppler factor $\delta = T_{\rm b} / T_{\rm int}$ becomes smaller than unity. If we assume a somewhat smaller $T_{\rm int} = 3 \times 10^{10}$~K proposed by \citet{Homan2006}, the Doppler factor is $\delta \approx 1$. In any case, the radio emission in J0839+5112 does not seem to be highly enhanced by relativistic beaming. It is possible that we underestimate the Doppler boosting by a factor of $\sim 2$ because the brightness temperature is measured at a high rest-frame frequency ($\sim 27$~GHz) where the intrinsic brightness temperature may be lower \citep{Lee2014}.   

\subsubsection{J2134$-$0419}

\begin{figure*}
\centering
\includegraphics[width=0.9\textwidth]{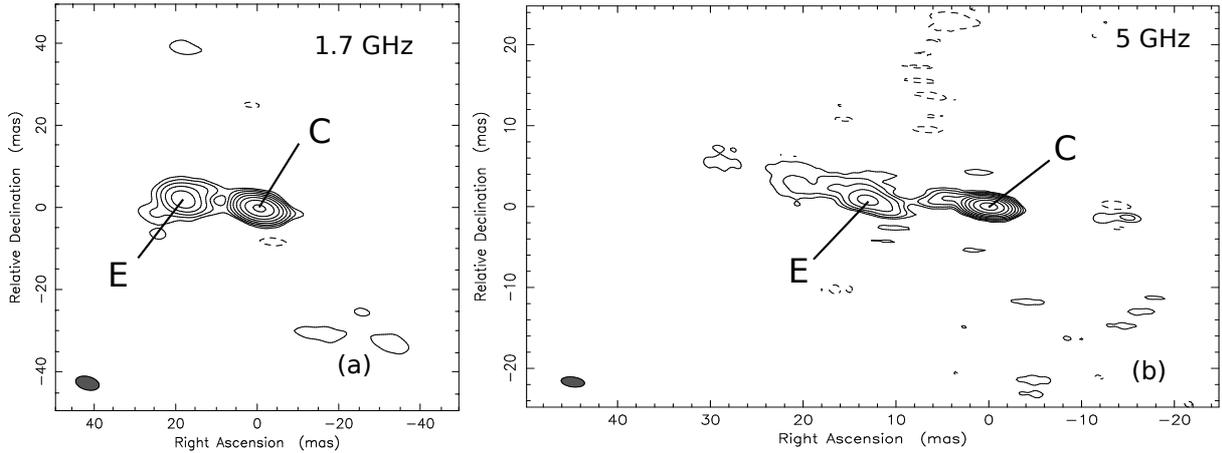}
\caption{Naturally weighted VLBI images of J2134$-$0419 at 1.7 and 5~GHz. 
The first contours are drawn at $\pm$3$\sigma$ image noise level. 
The positive contours increase by a factor of 2. The synthesized beam (FWHM) is shown at the lower-left corner of each image. The image parameters are listed in Table~\ref{tab:image-p}.}
\label{fig:2134}
\end{figure*}

The images produced from the phase-referenced and fringe-fitted visibility data show consistent structures at both frequencies. We present in Fig.~\ref{fig:2134} the images of J2134$-$0419 made after direct fringe-fitting which are based on more data and have higher dynamic range. The jet stucture is aligned to the east--west direction. Two components are seen in both the 1.7- and 5-GHz images, the westernmost one is the brighter and more compact. We identify this with the core. According to the accurate astrometry provided by the phase-referenced images, the core components positionally coincide at both frequencies. Two circular Gaussian model components were fitted to the brightness distribution of this source at 1.7 and 5~GHz, respectively (Table~\ref{tab:model-p}). The position of the eastern jet component (E) is slightly different at 5~GHz, consistent with its resolved nature apparent in Fig.~\ref{fig:2134}b, where the low surface brightness emission also traces out the jet structure towards the east.

The two-component structure of J2134$-$0419 is also seen in an image made with the Very Long Baseline Array (VLBA) at 4.3~GHz. This, and a 7.6-GHz image showing the core component with a marginally resolved eastern extension (within $\sim 2$\, mas) are availabe in the Astrogeo database\footnote{http://astrogeo.org/cgi-bin/imdb\_get\_source.csh?source=J2134-0419}. These short 1-min VLBA snapshots were taken after the acceptance of our EVN observing proposal, on 2015 September 1 (project code: BP192), about 2 and 6 weeks before our observations at 1.7 and 5~GHz, respectively. The overall core--jet structure towards the east is the same in the VLBA images and our more sensitive EVN images (Fig.~\ref{fig:2134}).     

In our EVN data at 1.7~GHz, the higher visibility amplitudes on the shortest Ef--Wb baseline indicate the presence of a large-scale structure which is resolved out with the EVN. Also, the image noise is significantly higher than the theoretical thermal noise level (21~$\mu$Jy\,beam$^{-1}$), suggesting some extended emission in the field. Indeed, the 1.4-GHz FIRST flux density (311.3~mJy, Table~\ref{tab:targets}) is $\sim$32 per cent higher than the integrated flux density recovered in our VLBI components (236.6~mJy, Table~\ref{tab:model-p}) at a close observing frequency of 1.7~GHz. This cannot be explained by spectral changes alone, and may be due to a radio structure extended to $\sim$0.1--1 arcsec scales, or/and flux density variability.

The dominant component of J2134$-$0419 has a flat spectrum ($\alpha = -0.2$). Its brightness temperature is $T_{\rm b} = (28.8 \pm 3.6) \times 10^{10}$~K, well exceeding $T_{\rm eq}$. The implied Doppler factor is $\delta \approx 6-10$. For the higher estimate, $T_{\rm int} = 3 \times 10^{10}$~K was assumed \citep{Homan2006}. The flat core spectrum and the large Doppler factor leave no doubt about the blazar identification of J2134$-$0419.

\subsubsection{J1420+1205}

\begin{figure*}
\centering
\includegraphics[width=0.9\textwidth]{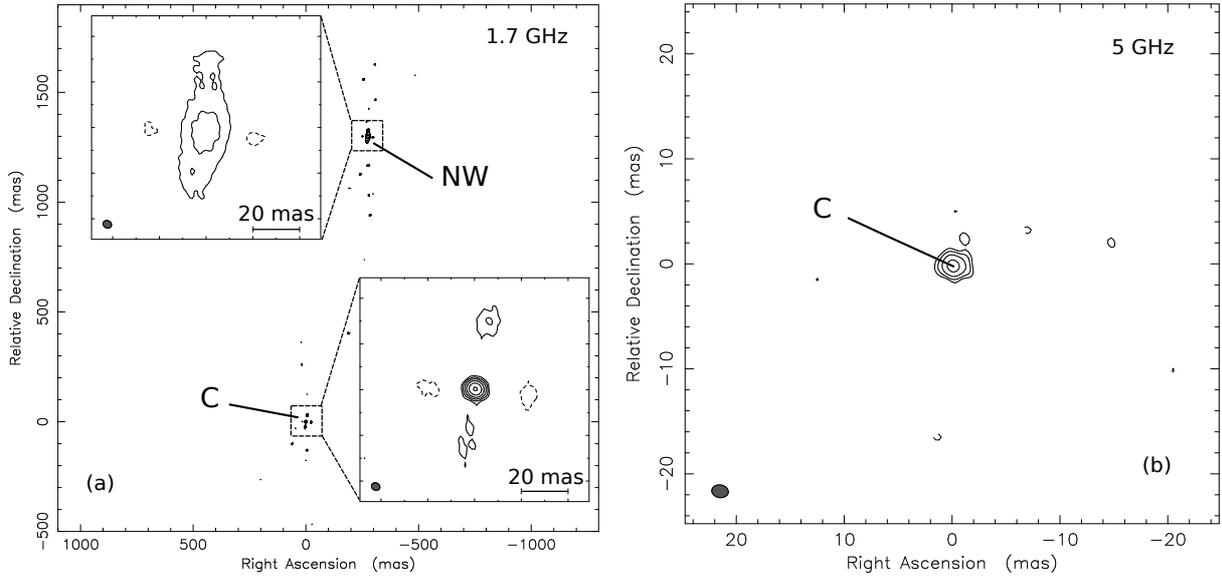}
\caption{Naturally weighted VLBI images of J1420+1205 at 1.7 and 5~GHz. 
At 5~GHz, only the south-eastern component was detected, which appears brighter and more compact in the 1.7-GHz image.
The first contours are drawn at $\pm$5$\sigma$ image noise level. 
The positive contours increase by a factor of 2. The synthesized beam (FWHM) is shown at the lower-left corner of each image. The image parameters are listed in Table~\ref{tab:image-p}.}
\label{fig:1420}
\end{figure*}

A wide double structure extended to $\sim1.33\arcsec$ is already apparent in the 1.7-GHz dirty image of this source. This size is smaller than the restoring beam of the VLA in the FIRST survey and thus the object appears unresolved in FIRST at 1.4~GHz. To restore the 1.7-GHz EVN image of J1420+1205 seen in Fig.~\ref{fig:1420}a, we fitted two circular Gaussian brightness distribution model components to the visibility data (Table~\ref{tab:model-p}). These represent the extendend emission better than {\sc clean} component models traditionally used in VLBI imaging. The brighter and more compact south-eastern component was also detected at 5~GHz (Fig.~\ref{fig:1420}b). One circular Gaussian model was fitted in this case, its brightness temperature is low, in the order of $10^{9}$~K (Table~\ref{tab:model-p}). The weaker, more extended north-western component seen in the 1.7-GHz image remained undetected at 5~GHz at the 6$\sigma$ image noise level of $\sim$0.5~mJy\,beam$^{-1}$.

The VLBI-recovered integrated flux density at 1.7~GHz (nearly 47.9~mJy, Table~\ref{tab:model-p}) is more than 45 per cent lower than the FIRST value at 1.4~GHz (Table~\ref{tab:targets}). This indicates that most of the radio emission originates from arcsec-scale extended structures in J1420+1205. The optical position of the object (right ascension $14^{\rm h}\,20^{\rm m}\,48\fs01$, declination $+12\degr\,05\arcmin\,45\farcs96$) from SDSS DR13 \citep{Albareti2016} coincides with the location of the more compact south-eastern VLBI component (Table~\ref{tab:position}) within the uncertainties. We can therefore identify this component with the galactic nucleus. The heavily resolved VLBI component to the north-west of the nucleus is most likely a lobe in a radio quasar, on the approaching side. The projected linear distance between the two components of J1420+1205 in the 1.7-GHz image is $\sim$9.5~kpc. In this picture, the apparent asymmetry of the structure may be caused by a viewing effect: the radio emission of the lobe on the receding side of the nucleus is below the detection threshold. Both the extended structure and the low brightness temperature of the central component indicate a large inclination angle of the structure with respect to the line of sight. This can hardly be reconciled with the blazar scenario.  

The lobe-dominated nature of J1420+1205 is consistent with its steep overall spectrum with $\alpha \approx -0.6$. The total flux densities of the source measured in a broad range of frequencies are available from the literature, at 74~MHz \citep[$750\pm130$~mJy,][]{Cohen2007}, 365~MHz \citep[$248\pm29$~mJy,][]{Douglas1996}, 1.4~GHz \citep[$92.1\pm2.8$~mJy,][]{Condon1998}, and 4.85~GHz \citep[$55\pm10$~mJy,][]{Gregory1991}.

\subsubsection{J2220+0025}

\begin{figure}
\centering
\includegraphics[width=0.40\textwidth]{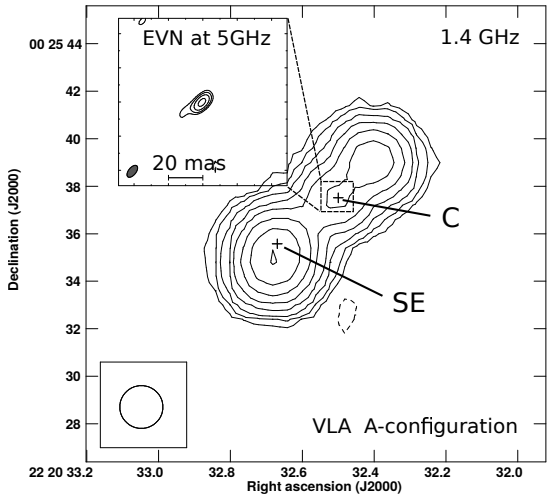}
\caption{The background image of J2220+0025 was made at 1.4~GHz with the VLA in its A configuration \citep{Hodge2011}. The circular Gaussian restoring beam size is $1.8\arcsec$ (FWHM), the first contours are drawn at $\pm$0.3~mJy\,beam$^{-1}$, the peak brightness is 40.05~mJy\,beam$^{-1}$. The positive contours increase by a factor of 2. The two + symbols indicate the positions of the two components seen 
in our 1.7-GHz EVN dirty map of J2220+0025 (not shown here). The 5-GHz EVN image of the only component detected is displayed in the inset. The data from baselines longer than 50 million wavelengths (50~M$\lambda$) were excluded. The first contours are drawn at $\pm$5$\sigma$ image noise level. The positive contours increase by a factor of 2. The image parameters are listed in Table~\ref{tab:image-p}.}
\label{fig:2220}
\end{figure}

This object falls in the sky area covered by the VLA--SDSS Stripe 82 survey \citep{Hodge2011}. This survey is about three times deeper than FIRST and mostly used the A configuration of the VLA that provides the longest baselines and thus the highest resolution. The 1.4-GHz VLA image of J2220+0025 reproduced in Fig.~\ref{fig:2220} shows an elongated structure extended to $\sim$10\arcsec. Imaging this extended source with VLBI proved difficult. At 1.7~GHz, there is a clear indication of two emission regions in the dirty image. Two circular Gaussian brightness distribution model components were fitted to the visibility data (Table~\ref{tab:model-p}). One of them coincides with the centre of the large-scale radio emission, the other one is very close to the brightness peak, located to the south-east of the centre. The sum of their flux densities (8.2~mJy) is much below the 1.4-GHz FIRST value (92.7~mJy, Table~\ref{tab:targets}), consistently with the resolved nature of the source.

Only the central component was clearly detected at 5~GHz with the high resolution of the EVN. We fitted a circular Gaussian model to characterise its properties (Table~\ref{tab:model-p}). The 5-GHz image of this faint but relatively compact feature is shown in the inset in Fig.~\ref{fig:2220}. This component is located in the SDSS DR 13 \citep{Albareti2016} optical position of the quasar (right ascension $22^{\rm h}\,20^{\rm m}\,32\fs49$, declination $+00\degr\,25\arcmin\,37\farcs49$) within the astrometric uncertainties. It is therefore right in the centre of a double-lobed radio AGN reminiscent of an FR-II type source \citep{Fanaroff1974} with a total projected linear extent up to $\sim$70~kpc. In this scenario, the south-eastern component detected with the EVN only at 1.7~GHz is a hot spot in the lobe on the approaching side.    

The VLA--SDSS Stripe 82 survey image (Fig.~\ref{fig:2220}) shows two major components. The south-eastern one is the brightest ($56.2\pm9.5$~mJy), the north-western component has $33.3\pm7.7$~mJy flux density at 1.4~GHz\footnote{http://sundog.stsci.edu/cgi-bin/searchstripe82}. The latter is a lobe blended with the weak core that we detected with the EVN. These two radio lobes add up the entire FIRST flux density ($92.7\pm6.4$~mJy, Table~\ref{tab:targets}) within the measurement errors. The 1.4-GHz flux density given in the NRAO VLA Sky Survey (NVSS) is $88.0 \pm 2.7$ \citep{Condon1998}, which also indicates the absence of radio structure on scales larger than $\sim$10$\arcsec$. If we assume that the structure of the double-lobed radio source is intrinsically symmetric, the flux density difference between the approaching jet side ($S_{\rm j}$) and the receding counterjet side ($S_{\rm cj}$) is caused by Doppler beaming. This allows us to estimate the viewing angle of the structure ($\vartheta$) with respect to the line of sight or the jet speed. According to e.g. \citet[][]{Arshakian2004}, the viewing angle of radio source with a continuous twin jet could be estimated as
\begin{equation}
\frac{S_{\rm j}}{S_{\rm cj}} = \left(\frac{1+\beta_{\rm j}\,\rm{cos}\vartheta}{1-\beta_{\rm j}\,\rm{cos}\vartheta} \right)^{2-\alpha}
\label{eq-4}
\end{equation}
where $\beta_{\rm j}$ is the jet speed expressed in the unit of light speed $c$. 
From a simple orientation-based unified model of radio AGN \citep{Barthel1989}, the viewing angle of quasar jets is expected at or below 45\degr. If we assume a typical value for spectral index $\alpha = -0.6$ \citep{Arshakian2004}, then we get $\beta_{\rm j} \approx 0.15$ for J2220+0025 at $\vartheta \approx 45\degr$.  Smaller inclination angles would infer even slower jet speed. Note that for low-redshift sources usually $\beta_{\rm j} = 0.4$ is assumed \citep{Arshakian2004}. In turn, higher $\beta_{\rm j}$ would result in inclination angles exceeding 45\degr, being inconsistent with the quasar classification \citep{Schneider2010}, unless there is substantial change in the orientation of the inner and outer jets. In any case, for reasonable $\beta_{\rm j}$ and $\alpha$ values, a robust result is that the inclination angle of the arcsec-scale structure is large in J2220+0025. 

We can conclude that, as in the case of J1420+1205, the compact radio emission decected with the EVN in this high-redshift source originates from an AGN as it shows brightness temperature well above the $\sim 10^{5}$~K limit for normal galaxies \citep{Condon1992}. Also, the monochromatic radio luminosity exceeds by orders of magnitude the lower limit of $\sim 2 \times 10^{21}$\,W\,Hz$^{-1}$ typical for radio AGN \citep{Kewley2010,Middelberg2011}. On the other hand, $T_{\rm b}$ is much lower than expected from a Doppler-boosted blazar jet. This, and the two-sided lobe-dominated structure inclined at a large angle to the line of sight seen in the VLA image are strong arguments against the blazar classification of J2220+0025.    

\section{Discussion}
\label{sec:Discussion}

\subsection{The compact radio sources J0839+5112 and J2134$-$0419}

The source J0839+5112 shows properties typical for blazars: compact mas-scale structure (Fig.~\ref{fig:0839}), flat spectrum and flux density variability. However, its moderate brightness temperature is close to the intrinsic value characteristic to a large sample of AGN in their low-brightness state \citep{Homan2006}, indidating no significant Doppler boosting. There is at least another radio quasar known at high redshift with similar properties. J1146+4037 at $z=5.01$ has a compact structure with moderate brightness temperature \citep{Frey2010,Coppejans2016} measured with VLBI but its X-ray data indicate its balazar nature \citep{Ghisellini2014}.  

Our EVN observations confirmed that J2134$-$0419 is a blazar. Moreover, the discovery of its prominent jet structure (Fig.~\ref{fig:2134}) will enable future attempts to detect apparent proper motion of the components. Like in the case of J1026+2542, a so far unique blazar at very high redshift ($z=5.27$) with directly estimated jet component proper motions \citep{Frey2015}, this may become feasible over a time baseline of 5--15 yr with multi-epoch VLBI measurements at 5~GHz. Our observations presented here could help detecting structural variations which appear slower by a factor of $(1+z)$ in the observer's frame because of the cosmological time dilation. In a separate study, we will attempt to compare the current VLBI data of J2134$-$0419 with archival data.

\subsection{Radio emission of the non-blazar sources J1420+1205 and J2220+0025}

We observed four blazar candidates with the EVN and found that, contrary to the expectations, two of them are not compact and are inclined at large angles to the line of sight. For J2220+0025 whose extended structure on arcsec scale was already known from VLA observations \citep{Hodge2011} we could identify a weak compact core in the centre of the large-scale radio structure, confirmig the double-lobed morphology. In these two sources, the radio emission is dominated by structures extended to $\sim 0.1-1$ arcsec scales. They also show clear evidence against Doppler-boosted inner (mas-scale) jet emission. Intriguingly, these two objects (J1420+1205 and J2220+0025) were claimed as the best blazar candidates in a complete sample of 19 extremely radio-loud quasars at $z>4$ by \citet{Sbarrato2015}, and X-ray observations with {\it Swift} seemed to confirm this notion, with estimated jet viewing angles of $\vartheta \approx 3\degr$.  

Recently \citet{Coppejans2016} collected data of all $z>4.5$ radio sources observed to date with VLBI. The selection criteria for these 30 objects partly taken from the literature are not homogeneous but most of them were targeted with VLBI because their radio emission is known to be compact in FIRST, i.e. with $\sim$5~arcsec angular resolution. Very similarly to the case of J1420+1205 and J2220+0025 presented in this paper, the source J1548+3335 ($z=4.68$) was found to have two widely separated ($\sim$0.8\arcsec, $\sim$5.3~kpc) components with the EVN at 1.7~GHz by \citet{Coppejans2016}. One of them was undetected at 5~GHz. The most plausible explanation of the radio properties of J1548+3335 is that a core (at both frequencies) and a hot spot (at 1.7~GHz only) in one of the two symmetric lobes in the kpc-scale extended structure of the source are seen with the EVN. This is our preferred explanation for J1420+1205 and J2220+0025 as well. Here the projected linear distances between the core and the brighter lobe are 9.5 and 21.5~kpc, respectively. Alternatively, the north-western component of J1420+1205 could also be interpreted as a foreground or background radio source physically unrelated to component C. Deep follow-up radio interferometric observations with $\sim 0.1$\,arcsec angular resolution could settle this question by detecting a symmetric lobe structure and possibly jet features connecting the core with the lobe(s). 

Another remarkable object in the list of 30 VLBI-imaged high-redshift radio AGN compiled by \citet{Coppejans2016} is J0311+0507 ($z=4.51$). A detailed radio interferometric study of this powerful FR-II source by \citet{Parijskij2014} found a complex two-sided core--jet--lobe stucture. The distance between the component identified with the galactic nucleus and the brighter hot spot in the south-western lobe is $\sim$1.3\arcsec (equals to about 8.7~kpc projected linear separation), comparable to what was measured in J1420+1205 and J1548+3335, and nearly 3 times smaller than in J2220+0025. The inclination angle of the structure with respect to the line of sight could be close to $45\degr$ in J0311+0507 \citep{Parijskij2014}.

Therefore our discovery of supposedly FR-II type arcsec-scale radio structures in the high-redshift AGN J1420+1205 and J2220+0025 is not unprecedented. In this scenario, the radio structures can naturally be interpreted as expanding young symmetric sources in the early Universe, as was also done for J0311+0507 by \citet{Parijskij2014}. Simply assuming a constant expansion speed $\beta_{\rm j} = 0.15$ and a moderate jet inclination $\vartheta$=45\degr, we obtain a rough estimate of their age, which is below $10^6$~yr. It is consistent with the value less than $10^6$~yr claimed for J0311+0507 by \citet{Parijskij2014}. What is puzzling is that both J1420+1205 and J2220+0025 show properties typical of blazars in X-rays \citep{Sbarrato2015}.

\subsection{X-ray emission of J1420+1205 and J2220+0025}

Since our VLBI data do not support the blazar classification of J1420+1205 and J2220+0025, the strong and hard X-ray emission and the overall SED \citep{Sbarrato2015} call for an alternative explanation. In general, AGN are prominent X-ray emitters \citep[e.g.][]{Rees1981,Schwartz2005}. For radio-quiet AGN, the X-rays are produced by inverse-Compton scattering of the UV/soft X-ray disk photons by the electrons in the hot corona. For radio-loud AGN, X-ray emission can also originate from the relativistic jet. In fact, the SED is dominated by the jet emission over the widest range of elecromagnetic radiation, from the radio to $\gamma$-rays. For blazars, the one-zone leptonic model \citep{Ghisellini2009} can successfully explain the double-humped SED. Here the dominant X-ray emission also results from inverse-Compton process, and comes from a small dissipation region in the jet near the central engine, where a high density of photons and relativistic electrons is expected.

Blazars are divided into three types according the position of the synchrotron peak frequency: low, intermediate, and high synchrotron-peak sources, abbreviated as LSPs, ISPs, and HSPs, respectively \citep[e.g.][]{Fan2016}. Different types may have different origin of seed photons  for the inverse-Compton hump \citep[e.g.][]{Kang2016}. The high-$z$ blazars belong to the LSPs, because they are very luminous, and luminosity appears to correlate with the frequency peak location \citep[the ``blazar sequence'',][]{Fossati1998,Ghisellini2016a}. Blazars often show strong $\gamma$-ray emission. However, there are no {\it Fermi} detections of $z>4$ blazars reported so far. For the very high redshift sources, the inverse-Compton peak moves into the X-ray band, and also the $\gamma$-rays are likely absorbed in the intergalactic space along the path from large cosmological distances. Only a couple of $\gamma$-ray blazars are identified at $3<z<4$ \citep[][]{Fan2016}. 

Can extended X-ray structures be responsible for the emission of J1420+1205 and J2220+0025? To date, kpc-scale X-ray jets have been found in more than 100 objects\footnote{https://hea-www.harvard.edu/XJET/}, the most distant of them is the blazar J1510+5702 at $z=4.3$ \citep{Siemiginowska2003,Cheung2004}. Their X-ray radiation mechanism is not well understood. Synchrotron emission of a secondary high-energy electron population is favored at low redshifts \citep[e.g.][]{Meyer2015}, but the acceleration mechanism producing the high-energy electrons is unclear. The inverse-Compton scattering of cosmic microwave background (CMB) photons by the relativistic electrons in the jet (the IC/CMB scenario) may also play an important role at high redshifts, where the energy density of CMB is $(1+z)^4$ times higher \citep[e.g.][]{Schwartz2002,Simionescu2016}. If the IC/CMB model works, then luminous X-ray jet components can even outshine the emission of the nucleus in high-$z$ sources. However, such sources are not found at $z>4$ so far \citep[e.g.][]{Wu2013}. The typical jet-to-core X-ray luminosity ratio is $\sim 0.01$ or smaller for lower-redshift quasars observed with {\it Chandra} \citep{Marshall2005}. 

The relativistic electrons in the radio lobes deposited by the jets cannot produce X-ray emission via synchrotron radiation because they have lower energies. At  redshifts above $z \sim 3$, the CMB energy density may dominate over the magnetic field energy density. This intensifies inverse-Compton cooling of the electrons, leading to an enhanced (diffuse) X-ray emission from the lobes \citep{Ghisellini2015b}. X-ray emission from radio lobes has been observed in several sources at redshift $z < 1$ and is interpreted as inverse-Compton scattering off CMB photons \citep[e.g.][]{Hardcastle2002}. However, there is no evidence for the diffuse nature of the X-ray emission in J1420+1205 and J2220+0025. On the contrary, \citet{Wu2013} found J1420+1205 unresolved with {\it Chandra} at a resolution of $\sim 1\arcsec$, which is however not much smaller than the overall size of the radio structure in our VLBI image. The X-ray source position measured from the image\footnote{Available from the Chandra archive, http://cda.harvard.edu/pop/} (right ascension $14^{\rm h}\,20^{\rm m}\,48\fs040$, declination $+12\degr\,05\arcmin\,46\farcs22$) is more consistent with that of the compact nuclear VLBI component in Fig.~\ref{fig:1420}a.

There are in principle two possibilities for the origin of X-ray emission in the two sources (J1420+1205 and J2220+0025) that are clearly not blazars based on our high-resolution VLBI imaging observations. However, none of the explanations is particularly compelling. 

\noindent {\it (1)} The X-ray emission may be a combination of the contribution from the (unbeamed) inner jet and from the kpc-scale jets and outer lobes (hot spots) of these expanding, powerful, presumably young radio sources. It is to be seen if any meaningful physical model can possibly reproduce the observed X-ray flux and spectral index, as well as the blazar-like SED of these AGN. 

\noindent {\it (2)} The X-rays originate from the jet launchnig region very close to the SMBH. The approaching jet then suddenly changes its direction on pc scale and appears already misaligned with respect to the line of sight and thus produces no Doppler-boosted radio emission on the scale probed by VLBI. While jet direction changes are not unprecedented at larger scales in radio AGN due to the interaction with the dense intestellar medium, it is unlikely to happen here, especially in two sources from a small sample of four.

Future sensitive spectral and high-resolution X-ray imaging observations of these sources, as well as the other two similar arcsec-scale extended high-$z$ sources, J0311+0507 \citep{Parijskij2014} and J1548+3335 \citep{Coppejans2016}, would help understand how these objects work and would provide the necessary input for detailed modeling of the X-ray emission. It is possible that other X-ray emitting, highly radio-loud AGN also mimic the typical blazar properties, yet they belong to different types of sources. This may contribute to the apparently larger fraction of blazars found in the early Universe \citep[cf.][]{Ghisellini2016b}. In the radio front, sensitive imaging at intermediate ($\sim$100-mas scale) resolution is essential to map the extended structure which is mostly resolved out with VLBI, to possibly reveal both of the symmetric lobes and the large-scale jets.   

\section{Conclusions}
\label{sec:Conclusions}

We performed EVN observations of four high-redshift ($z>4$) blazar candidates at 1.7 and 5~GHz frequencies. Compact radio emission from all four target sources were detected at both frequencies. Precise astrometric positions were determined with phase-referencing to nearby calibrator sources. However, the structures observed are spectacularly diverse in this small sample. The source J2134$-$0419 is clearly confirmed as a blazar with a flat-spectrum radio core and a Doppler-boosted jet emission. Its jet structure extending to $\sim$10~mas scale holds the promise of detecting jet component proper motion with multi-epoch 5-GHz VLBI imaging, a rare and valuable opportunity among high-redshift radio AGN. The source J0839+5112 has a compact mas-scale structure with flat spectrum, and an indication of flux density variability. Its brightness temperature is not enhanced by relativistic beaming which may indicate a larger inclination angle of the jet. 

Surprisingly, the two other blazar candidates, J1420+1205 and J2220+0025 show arcsec-scale extended structures, with most of their flux density resolved out on the long VLBI baselies. Their weak nuclear radio emission seems incompatible with the blazar scenario where the inner jet points nearly to the line of sight. Further sensitive interferometric observations at intermediate resolution would allow us to better characterise their radio structure. 

The fact that two of the candidates are clearly non-blazar sources requires the refinement of the broad-band SED modeling and poses an interesting question about the physical origin of their high-energy emission. Our results draw the attention to the utmost importance of VLBI imaging obervations for reliably classifying blazars at high redshift.

\section*{Acknowledgements}

The EVN is a joint facility of independent European, African, Asian, and North American 
radio astronomy institutes. Scientific results from data presented in this publication are derived from the following 
EVN project code: EC054.
H.-M. C. acknowledges support by the program of the Light in China's Western Region (Grant No. 2016-QNXZ-B-21 \& YBXM-2014-02) and 
the National Science Foundation of China (Grant No. 11573016 \& 11503072). 
S.F., K.\'E.G., and D.C. thank the Hungarian National Research, Development and Innovation Office (OTKA NN110333) for support.
T.A. thanks for the grant support by the Youth Innovation Promotion Association of the Chinese Academy of Sciences (CAS).
This work was supported by the China--Hungary Collaboration and Exchange Programme by the International Cooperation Bureau of the CAS
and the Program for Innovative Research Team (in Science and Technology) at the University of Henan Province.
We acknowledge the use of Astrogeo Center database of brightness distributions, correlated flux densities, and images of compact radio sources produced with VLBI.
This research has made use of the NASA/IPAC Extragalactic Database (NED) which is operated by the Jet Propulsion Laboratory, California Institute of Technology, under contract with the National Aeronautics and Space Administration. 






\bsp	
\label{lastpage}
\end{document}